\documentclass[journal]{IEEEtran}

\ifCLASSINFOpdf
\else
   \usepackage[dvips]{graphicx}
\fi
\usepackage{url}

\usepackage{graphicx}
\usepackage{hyperref}
\usepackage{multirow}
\usepackage{booktabs} 
\usepackage{amsmath}
\usepackage{amssymb}
\usepackage{orcidlink}
\usepackage{subcaption}  
\usepackage{caption} 
\usepackage{float}
\usepackage{subcaption}
\usepackage[table]{xcolor}
\usepackage{pifont} 
\newcommand{\cmark}{\ding{51}}
\newcommand{\xmark}{\ding{55}}

\begin{document}

\title{Cross-Dialect Bird Species Recognition with Dialect-Calibrated Augmentation}

\author{Jiani Ding\orcidlink{0009-0004-7706-6849}, Qiyang Sun\orcidlink{0009-0001-9228-4543}, Alican Akman\orcidlink{0000-0002-8010-6897}, and Björn W. Schuller\orcidlink{0000-0002-6478-8699} \IEEEmembership{Fellow, IEEE}

\thanks{Jiani Ding, Qiyang Sun and Alican Akman are with GLAM, Department of Computing, Imperial College London, UK (e-mail: jiani.ding24@imperial.ac.uk; q.sun23@imperial.ac.uk; a.akman21@imperial.ac.uk).}

\thanks{Björn W. Schuller is with GLAM, Department of Computing, Imperial College London, UK; CHI - Chair of Health Informatics, TUM University Hospital, Munich, Germany; MDSI – Munich Data Science Institute, Munich, Germany; and MCML – Munich Center for Machine Learning, Munich, Germany (e-mail: bjoern.schuller@imperial.ac.uk).}

\thanks{Jiani Ding and Qiyang Sun contributed equally to this work.}

}
\markboth{Journal of \LaTeX\ Class Files, Vol. 14, No. 8, August 2015}
{Shell \MakeLowercase{\textit{et al.}}: Bare Demo of IEEEtran.cls for IEEE Journals}
\maketitle

\begin{abstract}
Dialect variation hampers automatic recognition of bird calls collected by passive acoustic monitoring. We address the problem on DB3V, a three-region, ten-species corpus of 8-s clips, and propose a deployable framework built on Time-Delay Neural Networks (TDNNs). Frequency-sensitive normalisation (Instance Frequency Normalisation and a gated Relaxed-IFN) is paired with gradient-reversal adversarial training to learn region-invariant embeddings. A multi-level augmentation scheme combines waveform perturbations, Mixup for rare classes, and CycleGAN transfer that synthesises Region 2 (Interior Plains)-style audio, , with Dialect-Calibrated Augmentation (DCA) softly down-weighting synthetic samples to limit artifacts. The complete system lifts cross-dialect accuracy by up to twenty percentage points over baseline TDNNs while preserving in-region performance. Grad-CAM and LIME analyses show that robust models concentrate on stable harmonic bands, providing ecologically meaningful explanations. The study demonstrates that lightweight, transparent, and dialect-resilient bird-sound recognition is attainable.
\end{abstract}


\IEEEpeerreviewmaketitle

\section{Introduction}

\IEEEPARstart{B}{ird} vocalisations are vital proxies of ecosystem health; yet, the sheer volume of audio gathered by passive acoustic monitoring (PAM) makes manual annotation infeasible \cite{eldesoky2025bird}. Recent deep-learning systems can identify bird species automatically, but their field utility is curbed by three intertwined barriers: (i) dialectal domain shift: the same species sings differently across regions, causing steep accuracy drops when a model is deployed outside its training geography; (ii) limited compute on edge devices that must run continuously in remote habitats; and (iii) the need for interpretable outputs that ecologists can trust when formulating conservation policy \cite{heinrich2025audioprotopnet}.

Culturally transmitted `dialects' can diverge within a species over only few distances, shaping territory defence and mate choice \cite{podos2007evolution}.  
Large-scale PAM studies report that recognisers fail unless dialect shift is addressed \cite{terrigeol2022efficiency, soanes2023passive}.  
Quantitatively, accuracy of a TDNN trained in one North-American region drops by up to 40 \% on another \cite{xie2023cross}, echoing the cross-site gaps reviewed in computational bioacoustics \cite{stowell2022computational}.
\begin{figure*}[htbp]
    \centering
    \includegraphics[width=1\linewidth]{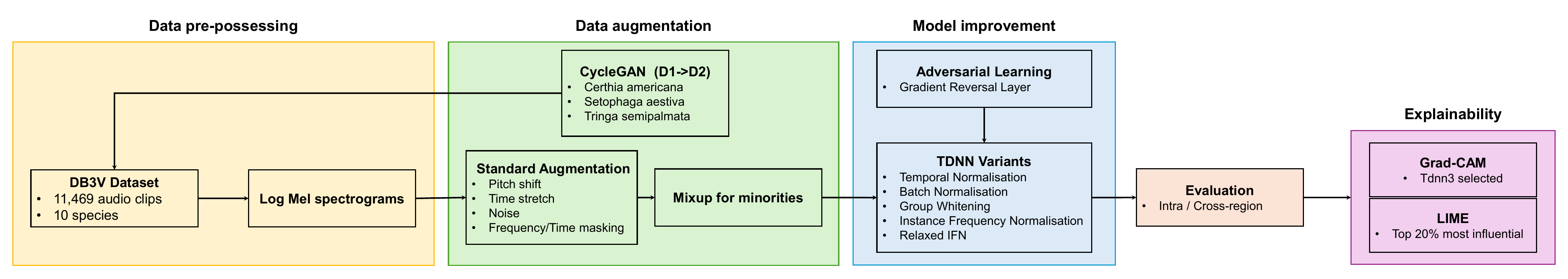}
    \caption{Workflow of the whole structure}
    \label{fig:workflow}
\end{figure*}
Domain-generalisation tactics range from heavy transformers to feature normalisation.  
Instance-Frequency Normalisation (IFN) wins multi-device benchmarks by discarding channel bias, yet, keeping species' cues \cite{kim2022domain}.  
Compact back-bones such as MobileNet, TDNN x-vectors, and depth-wise CNNs offer $8\times$ fewer MACs than ResNet while matching AudioSet accuracy \cite{sandler2018mobilenetv2}. 
Pre-training on AudioSet then fine-tuning for birdsong further boosts low-resource performance \cite{kong2020panns}, but dialect-centric evaluation is still rare.  
Prior work on audio domain adaptation has improved cross-region accuracy, yet, typically relies on heavy transformer back-bones \cite{mu2023improving} and offers little transparency into decision making.  
For bioacoustic applications, however, models must be compact, energy-efficient and explainable \cite{gascoyne2025first, silva2025impacts}.

Front-end choices and normalisation also matter: PCEN \cite{lostanlen2018per} and learnable front-ends like LEAF \cite{zeghidour2021leaf} improve robustness under device and distance mismatch, while IFN/RIFN explicitly target frequency-wise channel bias; these components are attractive for edge-constrained deployments. 

Early bioacoustic studies relied on pitch-shift, time-stretch, and additive noise to enlarge training corpora and improve noise robustness \cite{salamon2017deep}.  
SpecAugment later showed that simple time- and frequency-masking can boost CNN and TDNN performance at virtually no extra cost \cite{park2019specaugment}.  
To combat severe class imbalance, Mixup interpolates minority-class spectrograms, smoothing decision boundaries and reducing over-fitting \cite{zhang2018mixup}.  
More recently, generative models have been explored: CycleGAN-VC architecture transfers speaker timbre in speech \cite{kaneko2019cyclegan, sun2024audio}, and similar voice-conversion GANs have been used to translate between recording devices or habitats, yielding modest cross-site gains \cite{tan2021gan}.
In parallel, community benchmarks such as BirdCLEF emphasise geographically diverse soundscapes and repeatedly report strong location effects, underscoring the need to evaluate augmentation and adaptation specifically for cross-region transfer \cite{kahl2021overview, kahl2022overview}.

Post-hoc explainability tools such as Grad-CAM, SHAP, and LIME have been ported to spectrograms, exposing frequency regions implicated in the prediction \cite{sun2025explainable,li2024detecting, lundberg2017unified}.  
Das et al.\ compared these maps under noise and adversarial perturbation for bird-CNNs \cite{das2024exploring}.  
Yet, saliency behaviour under dialect shift remains unexplored. 
Recent explainable architectures (e.g., prototype-based audio models) and embedding visualisation further illustrate how explanation can inform model trust, but they have rarely been assessed under regional domain shift \cite{heinrich2025audioprotopnet, ghani2023global}.
No previous work jointly tackles (i) compact modelling, (ii) dialect robustness, and (iii) visual explainability.

The DB3V benchmark \cite{jing2024db3v} addresses this gap by providing ten North-American species recorded in three geographically distinct regions, each exhibiting measurable dialect variation, and therefore serves both as a realistic test-bed for dialect-robust recognition and as a proxy for deployment-level constraints.

This work makes three contributions:

\begin{itemize}
    \item We propose a lightweight TDNN enhanced with instance-frequency normalisation and an adaptive gradient-reversal layer, delivering strong cross-region accuracy at a fraction of the compute cost of typical large models.
    \item We achieve a new state-of-the-art (SOTA) on the DB3V corpus, improving unseen-dialect accuracy by 11\% via Dialect-Calibrated Augmentation (DCA) that integrates conventional perturbations, Mixup for rare species, and CycleGAN-based spectrogram transfer.
    \item We adapt Grad-CAM and LIME to Mel-spectrograms, revealing which spectral cues drive predictions and validating model reliability under dialectal shift.
\end{itemize}
This report is organised as follows: Section \ref{sec:method} outlines the task, network design, normalisation options, and the two-stage augmentation strategy.
 Section \ref{sec:result} describes the experimental setup, covering the DB3V corpus, training protocol, and evaluation metrics, and then presents the results, ablation studies, and saliency analyses that trace how each component improves cross-dialect accuracy.
 Section \ref{sec:conclusion} concludes with key findings and future works.

\section{Proposed Method}
\label{sec:method}
\begin{figure}
    \centering
    \includegraphics[width=1\linewidth]{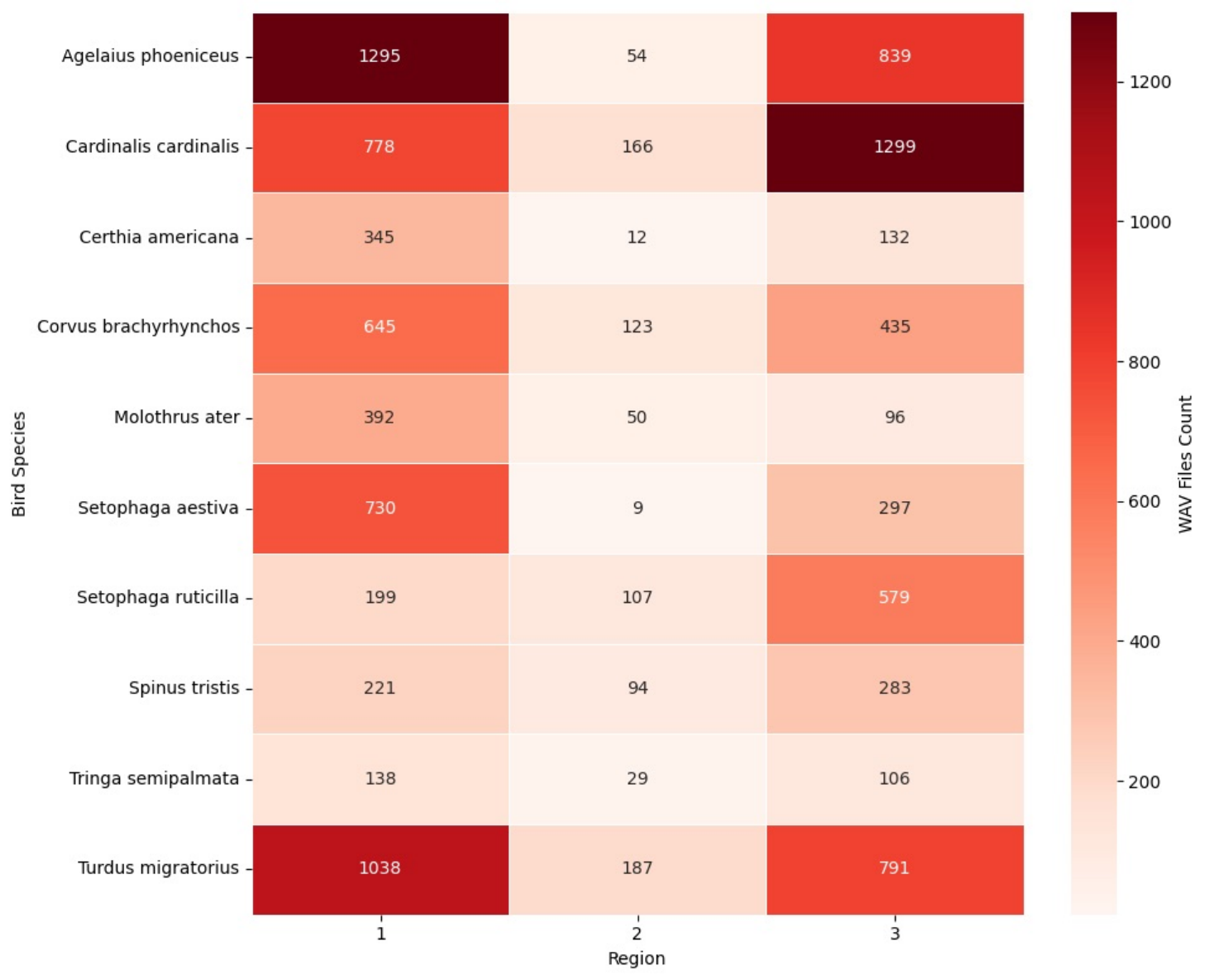} 
    \caption{Distribution of bird species audio recordings across three regions.}
    \label{fig:heatmap}
\end{figure}

\subsection{Task Definition}
An 8-s waveform $x$ is transformed into a $F{\times}T=128{\times}251$ log-Mel spectrogram $\mathbf{X}$.  
The model learns $f_{\theta}\!:\mathbf{X}\!\mapsto\!y,\;y{\in}[1,10]$ while discarding region bias $d\,{\in}\,[1,3]$ available only at training.  
We then present DCA as one-line scheduler that down-weights synthetic dialect-transfer examples to limit artifacts while preserving diversity. In this paper we instantiate DCA with a fixed w\_{\text{syn}} = 0.3, yielding the reported improvements without any additional tuning.
The training objective is to minimize:

\begin{equation}
\label{eq:loss}
\mathcal{L} = \frac{1}{N} \sum_{i=1}^N \left( w_i \mathcal{L}_{CE}^{\text{sp}}(y_i | x_i) + \alpha \mathcal{L}_{CE}^{\text{dom}}(d_i | x_i^{\text{GRL}}) \right).
\end{equation}

Samples produced by CycleGAN are down-weighted to \(w_{\text{syn}}=0.3\), while all other samples keep unit weight. We set \(\alpha=0.5\) and apply the domain term via a Gradient-Reversal Layer (GRL) \cite{ganin2015unsupervised}.
During training, the layer multiplies the backward gradient by a factor that is linearly warmed up from 0.1 to 1.0 over the first ten epochs, so the dialect-invariance constraint is introduced gradually.


\begin{table*}[htbp]
\centering
\caption{Baseline model performance on intra-region and cross-region evaluations (part)}
\label{tab:baseline_all_regions}
\begin{tabular}{ll|ccc|ccc|ccc}
\toprule
 & & \multicolumn{3}{c|}{\textbf{Test: D1}} & \multicolumn{3}{c|}{\textbf{Test: D2}} & \multicolumn{3}{c}{\textbf{Test: D3}} \\
\textbf{Model} & \textbf{Train} & ACC & UAR & F1 & ACC & UAR & F1 & ACC & UAR & F1 \\
\midrule
\multirow{3}{*}{TDNN+BN}
  & D1 & 0.9655 & 0.9629 & 0.9634 & 0.5952 & 0.4660 & 0.4248 & 0.5861 & 0.5304 & 0.4996 \\
  & D2 & 0.2707 & 0.2949 & 0.2156 & 0.8929 & 0.8813 & 0.8659 & 0.3852 & 0.3001 & 0.2835 \\
  & D3 & 0.5948 & 0.5301 & 0.5088 & 0.7500 & 0.6366 & 0.5662 & 0.9631 & 0.9403 & 0.9529 \\
\midrule
\multirow{3}{*}{TDNN+GW}
  & D1 & 0.9500 & 0.9402 & 0.9422 & 0.5833 & 0.5025 & 0.4173 & 0.6127 & 0.5627 & 0.5381 \\
  & D2 & 0.2983 & 0.3307 & 0.2466 & 0.9167 & 0.7541 & 0.7655 & 0.4631 & 0.3730 & 0.3535 \\
  & D3 & 0.5948 & 0.5505 & 0.5240 & 0.7619 & 0.6145 & 0.5844 & 0.9488 & 0.9119 & 0.9190 \\
\midrule
\multirow{3}{*}{TDNN+IFN}
  & D1 & 0.8983 & 0.8800 & 0.8905 & 0.7500 & 0.7037 & 0.6285 & \textbf{0.6926} & 0.6267 & 0.6008 \\
  & D2 & 0.3724 & 0.3685 & 0.2731 & 0.7857 & 0.6176 & 0.6277 & 0.5020 & 0.3987 & 0.3471 \\
  & D3 & \textbf{0.6586} & \textbf{0.5927} & \textbf{0.5640} & \textbf{0.8690} & \textbf{0.8270} & \textbf{0.8420} & 0.8668 & 0.7788 & 0.7987 \\
\midrule
\multirow{3}{*}{TDNN+RIFN}
  & D1 & 0.8948 & 0.8829 & 0.8858 & 0.6548 & 0.6431 & 0.5742 & 0.6865 & \textbf{0.6584} & \textbf{0.6332} \\
  & D2 & 0.4345 & 0.4406 & 0.3646 & 0.8214 & 0.7060 & 0.7019 & 0.5881 & 0.4556 & 0.4241 \\
  & D3 & 0.6414 & \textbf{0.5943} & \textbf{0.5664} & 0.8095 & 0.7495 & 0.7166 & 0.8668 & 0.7957 & 0.8103 \\
\midrule
\multirow{3}{*}{TDNN+TN}
  & D1 & 0.9672 & 0.9635 & 0.9642 & 0.4881 & 0.3307 & 0.3008 & 0.4447 & 0.3905 & 0.3609 \\
  & D2 & 0.2362 & 0.2736 & 0.1975 & 0.9286 & 0.7723 & 0.7857 & 0.4098 & 0.2986 & 0.2843 \\
  & D3 & 0.5121 & 0.4692 & 0.4375 & 0.6429 & 0.5552 & 0.4714 & 0.9611 & 0.9303 & 0.9440 \\
\bottomrule

\end{tabular}

\end{table*}

\subsection{Normalisation Strategies}
Waveforms are resampled to 16 kHz; STFT (2 048 FFT / 512 hop) plus 128 Mel filters yield $\mathbf{X}$.  
Consistent with the previous study \cite{jing2024db3v}, we explore 5 normalisers: Batch Normalisation 
\cite{ioffe2015batch}, Group Whitening \cite{huang2021group}, Time-Norm \cite{xie2023cross}, IFN, and Relaxed-IFN \cite{xie2023cross, kim2022domain}.

\subsection{Data Augmentation}
\noindent\textbf{CycleGAN pre-expansion.}  
Region-2 (Interior Plains) minority species are augmented by style-transferring Region-1 (Western Cordillera) recordings with CycleGAN-VC2 \cite{kaneko2019cyclegan}.  
Synthetic clips are down-weighted when computing Eq.~\eqref{eq:loss}. 

\noindent\textbf{In-training pipeline.}  
For all regions, we apply pitch/time shifts, Gaussian noise, SpecAugment masks \cite{park2019specaugment}, and Mixup \cite{zhang2018mixup}.  
Class-aware sampling equalises minority presence per mini-batch.

\subsection{Explainability}
To validate ecological plausibility, two complementary post-hoc tools are applied on the frozen model:  
(i) Grad-CAM highlights class-discriminative time–frequency regions \cite{selvaraju2017grad}; 
(ii) LIME perturbs individual spectrogram tiles to yield model-agnostic local attributions \cite{ribeiro2016should}.  Comparing saliency distributions across dialects pinpoints whether performance drops stem from shifted frequency bands or spurious artifacts, closing the loop between robustness and interpretability.

\section{Experimental Results and Analysis}
\label{sec:result}
\subsection{Dataset and Protocol}
We use the DB3V corpus \cite{jing2024db3v}: ten North-American species (11469 clips) recorded in three dialectal regions~(D1–D3).  Figure~\ref{fig:heatmap} visualises the species–region distribution, highlighting the
severe scarcity of several species in D2.

Models are trained on one region and evaluated on all three, yielding nine train$\!\rightarrow$test pairs.  

We report Accuracy, macro-F1 and, Unweighted Average Recall (UAR), averaged over five repeated  experiments.  Class-imbalance is mitigated by weighted sampling. Fig. \ref{fig:workflow} illustrates the whole workflow of the study. Code is available on our project page \footnote{\url{https://github.com/JianiD/bird_vocalization_db3v_improvement}}.

\subsection{Baseline Model Comparison}
Consistent with  Jing et al. \cite{jing2024db3v}, all models achieve near-ceiling accuracy when training and testing on the same region. However, performance drops sharply under dialectal shift, reflecting the well-documented difficulty of dialect transfer (Table~\ref{tab:baseline_all_regions}).

Across all architectures, Region D2 consistently emerges as the weakest evaluation domain. Models trained on D2 generalise particularly poorly to other regions. Even in the in-region setting, training and testing on D2 yields substantially lower performance than equivalent experiments on D1 or D3. This confirms the dataset imbalance, where D2 is severely underrepresented. 

Among normalisation strategies, IFN consistently outperforms other models in cross-region tasks and therefore serves as our strong baseline. RIFN also demonstrates robustness improvements, though its within-dialect performance is marginally lower than IFN.

\subsection{Augmentation and Adversarial Learning}
\label{sec:aug_adv}
Starting from the TDNN + IFN baseline in
Table~\ref{tab:baseline_all_regions}, we introduce DCA and GRL in a step-wise fashion:  
(i)~\emph{perturbation-based augmentation}
(waveform and spectrogram);  
(ii)~\emph{adv.\ domain confusion} via a gradient-reversal layer (GRL);  
(iii)~\emph{Mixup} for minority classes; and  
(iv)~\emph{CycleGAN style transfer}, which converts Region-1 clips to the acoustic style of Region-2 (with DCA weight w\_{\text{syn}}=0.3).

\noindent\textbf{Ablation study.}
Table~\ref{tab:ablation_ifn_acc} reports the step-wise ACC when training on the lowest-resource dialect~D2.  
Standard perturbations increase in-domain accuracy slightly, yet, hurt cross-dialect generalisation.  
Adding GRL yields a clear boost on D1/D3, confirming the benefit of representation alignment.  
Mixup offers marginal extra gains, while the largest jump comes from CycleGAN augmentation.

\begin{table}[htbp]
\centering
\caption{Ablation on TDNN+IFN trained on D2 (ACC \%).}
\label{tab:ablation_ifn_acc}
\begin{tabular}{lccc}
\toprule
\textbf{Variant} & \textbf{Test D1} & \textbf{Test D2} & \textbf{Test D3} \\
\midrule
Baseline            & 37.24 & 78.57 & 50.20 \\
\ + Standard Aug.        & 36.72 & 80.95 & 47.13 \\
\ + Adv.\ (GRL)          & 50.34 & 75.86 & 62.24 \\
\ + Mixup                & 51.03 & 79.31 & 62.45 \\
\ + CycleGAN$^{\dagger}$ & 45.36 & \textbf{87.36} & \textbf{63.88} \\
\bottomrule
\end{tabular}

\end{table}

\begin{figure}
    \centering
    \includegraphics[width=1\linewidth]{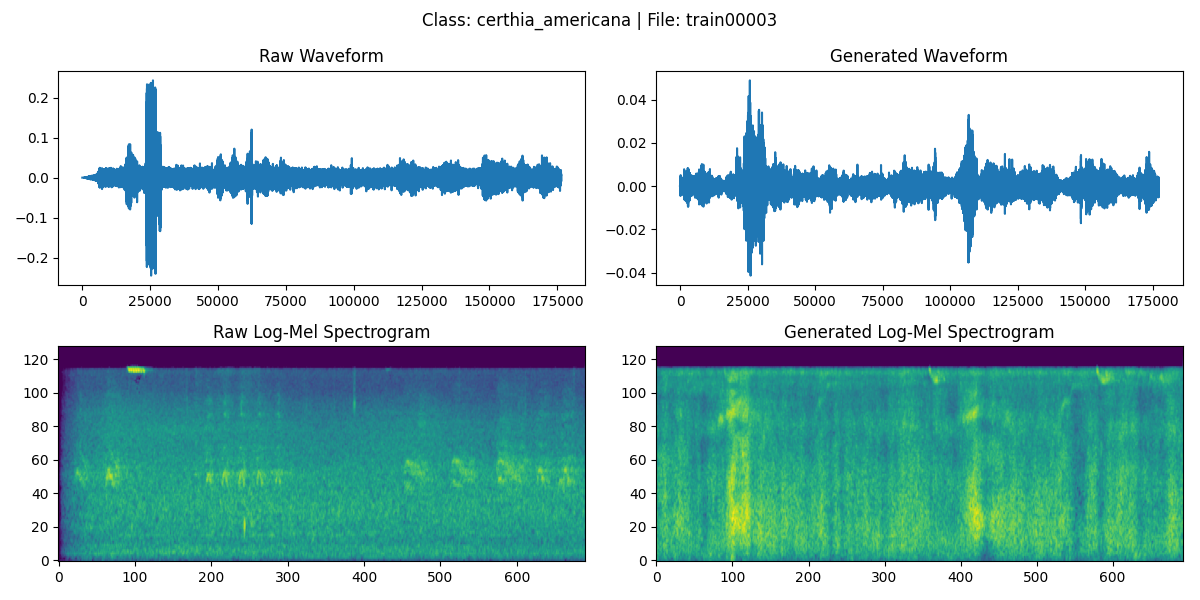}
    \caption{Real log-Mel vs. CycleGAN spectrograms.}
    \label{fig:cyclegan_ca}

\end{figure}

\noindent\textbf{Qualitative Evaluation of CycleGAN Outputs.}
Before integrating generated samples into training, we conducted a qualitative inspection of the spectrograms produced by the CycleGAN-VC2 model. 
Figure~\ref{fig:cyclegan_ca} compares real and
generated examples for \textit{Certhia americana}, an under-represented species in Region 2.

The synthetic samples closely mimic the global harmonic structure and temporal envelope of the originals. While minor distortions are visible in part bands, the core syllabic patterns and frequency contours remain intact. This suggests that the generator successfully preserves ecologically salient features while altering regional style.

However, given the absence of ground-truth alignment, we avoid over-reliance on synthetic data. As a safeguard, all CycleGAN-generated samples were integrated into training with a reduced DCA loss weight 0.3.

\noindent\textbf{Final model vs.\ baseline.}
Table~\ref{tab:acc_compact} reports the effect of combining all techniques which yields our fully-enhanced TDNN+IFN.  
For the lowest-resource Region D2, TDNN+IFN accuracy rose relatively 11.2\% for intra-dialect test, and achieved 21.8\% and 27.3\% increases for cross-domain tests. These results validate our multi-level strategy: frequency-aware normalisation, diversity-oriented augmentation, and adversarial alignment must act together to achieve robust cross-dialect recognition.

\begin{table}[htbp]
\centering
\caption{Final TDNN+IFN vs. baseline (ACC \%, $\Delta$)}
\label{tab:acc_compact}
\setlength{\tabcolsep}{4pt}
\renewcommand{\arraystretch}{1.1}
\begin{tabular}{c|cc|cc|cc}
\toprule
\multirow{2}{*}{\textbf{Train}} &
\multicolumn{2}{c|}{\textbf{Test D1}} &
\multicolumn{2}{c|}{\textbf{Test D2}} &
\multicolumn{2}{c}{\textbf{Test D3}} \\
\cmidrule(lr){2-3}\cmidrule(lr){4-5}\cmidrule(l){6-7}
 & ACC & $\Delta$ & ACC & $\Delta$ & ACC & $\Delta$ \\
\midrule
D1 & 90.21 & +0.4 & 74.71 & –0.4 & 71.02 & +2.5 \\
D2 & 45.36 & +21.8 & 87.36 & +11.2 & 63.88 & +27.3 \\
D3 & 66.32 & +0.7 & 81.61 & –6.1 & 89.39 & +3.1 \\
\bottomrule
\end{tabular}
\end{table}

\noindent\textbf{Analysis.}
Normalisation + data-level adaptation + GRL yields the best overall robustness. Neither frequency-aware normalisation nor augmentation alone suffices; their combination, supplemented by representation-level domain confusion, produces the highest and most consistent scores, especially in the under-represented Region D2.
These findings validate our hypothesis that robust cross-dialect classification requires a multi-level strategy: architecture-aware normalisation to suppress spurious channel cues, targeted style transfer to enrich low-resource dialects, and adversarial learning to align latent representations.

\subsection{Explainability}

We also probe model attention with Grad-CAM and LIME.  
All examples in this section use audio samples from Region 1 so that differences arise solely from training dialect.  

\begin{figure}[t]
\centering
\begin{subfigure}{.49\linewidth}
    \includegraphics[width=\linewidth]{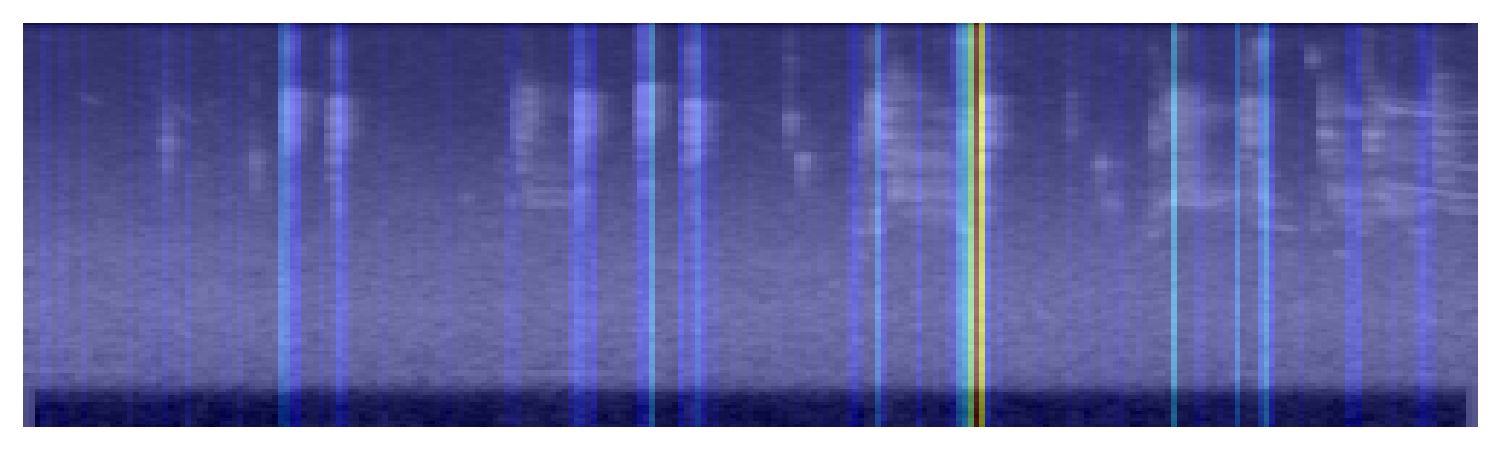}
    \includegraphics[width=\linewidth]{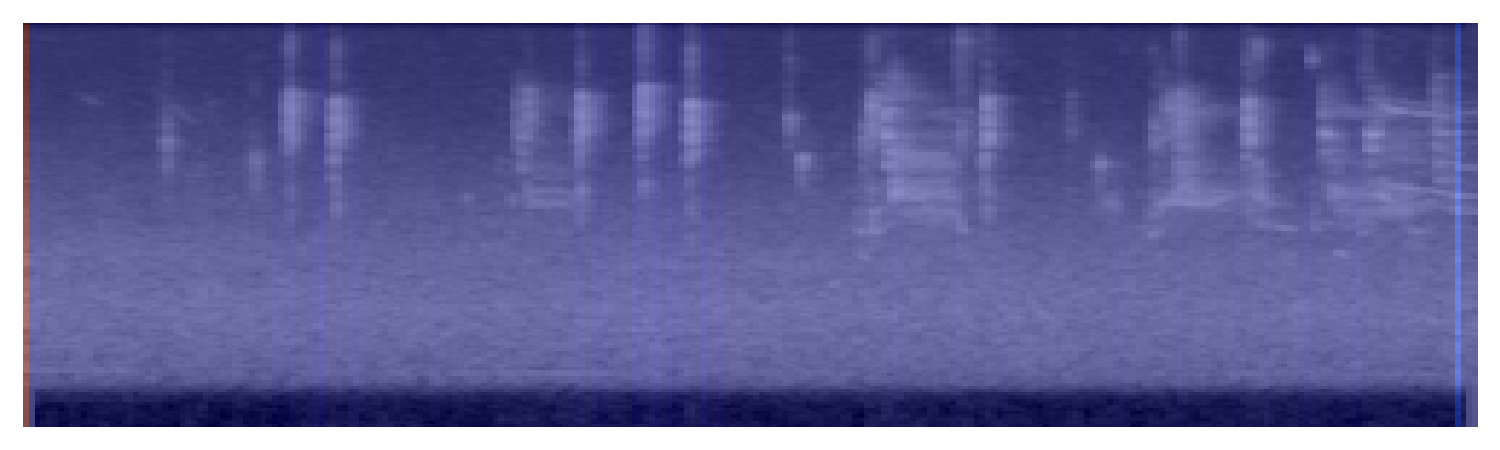}
    \includegraphics[width=\linewidth]{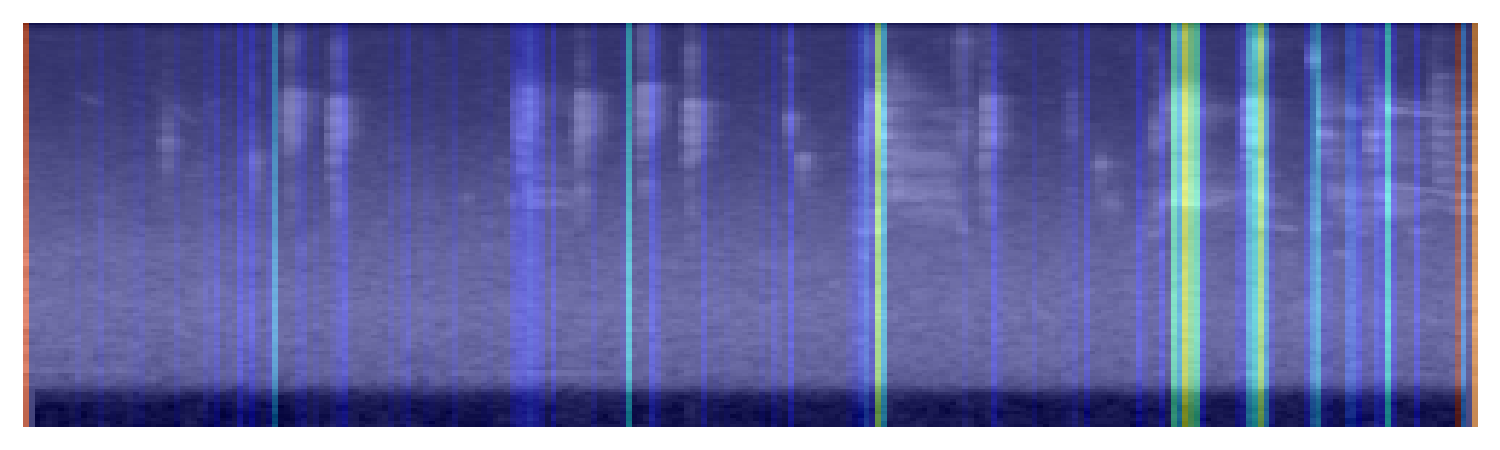}
    \caption*{Grad-CAM (rows: \cmark, \xmark, \xmark)}
\end{subfigure}%
\hfill
\begin{subfigure}{.49\linewidth}
    \includegraphics[width=\linewidth]{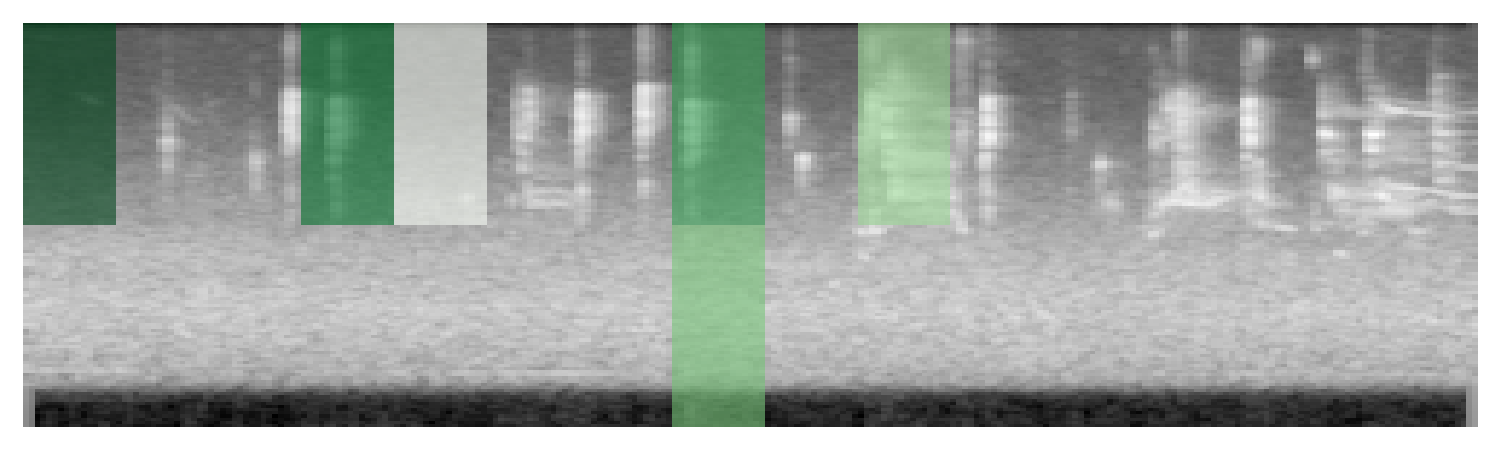}
    \includegraphics[width=\linewidth]{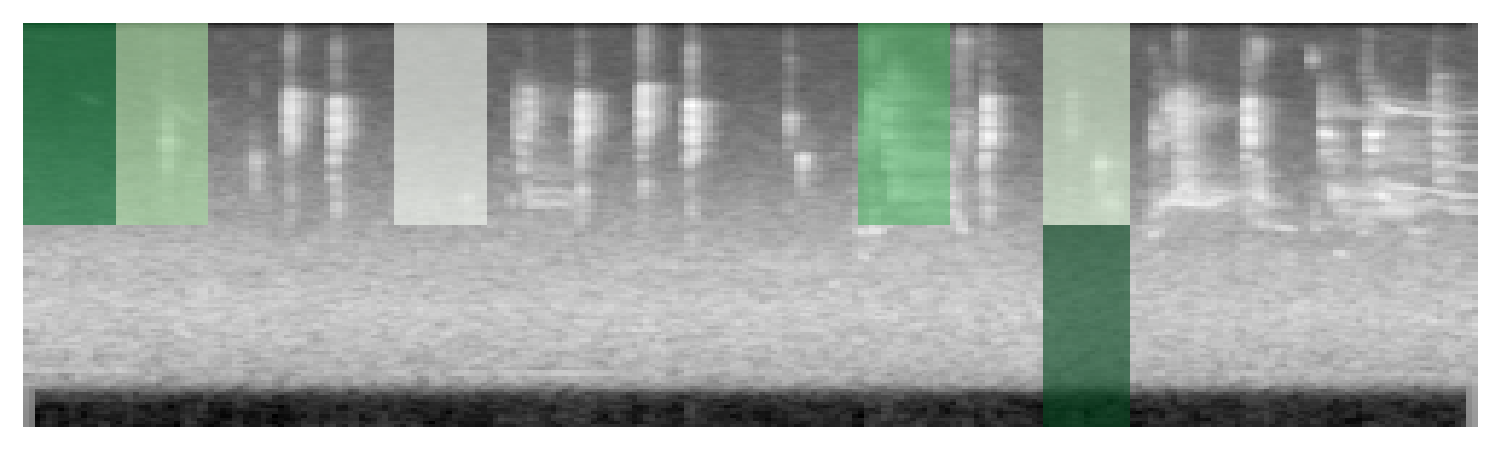}
    \includegraphics[width=\linewidth]{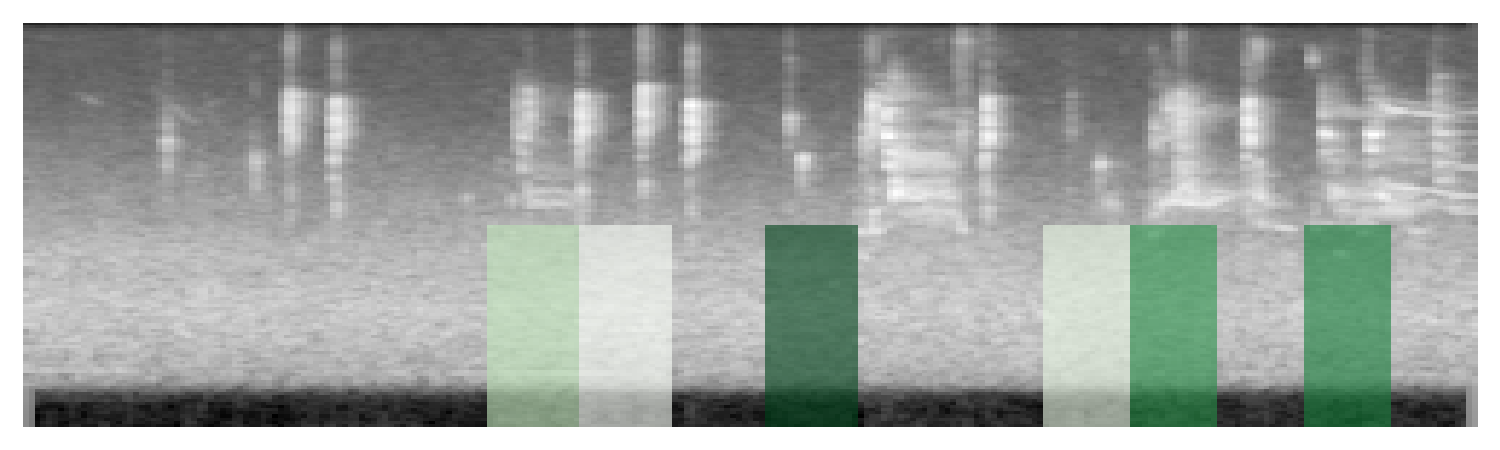}
    \caption*{LIME (rows: \cmark, \xmark, \xmark)}
\end{subfigure}

\caption{Misclassified sample (\textit{A.\,phoeniceus}). Rows: models trained on D1, D2, D3.}
\label{fig:case1}
\end{figure}

\begin{figure}[t]
\centering
\begin{subfigure}{.49\linewidth}
    \includegraphics[width=\linewidth]{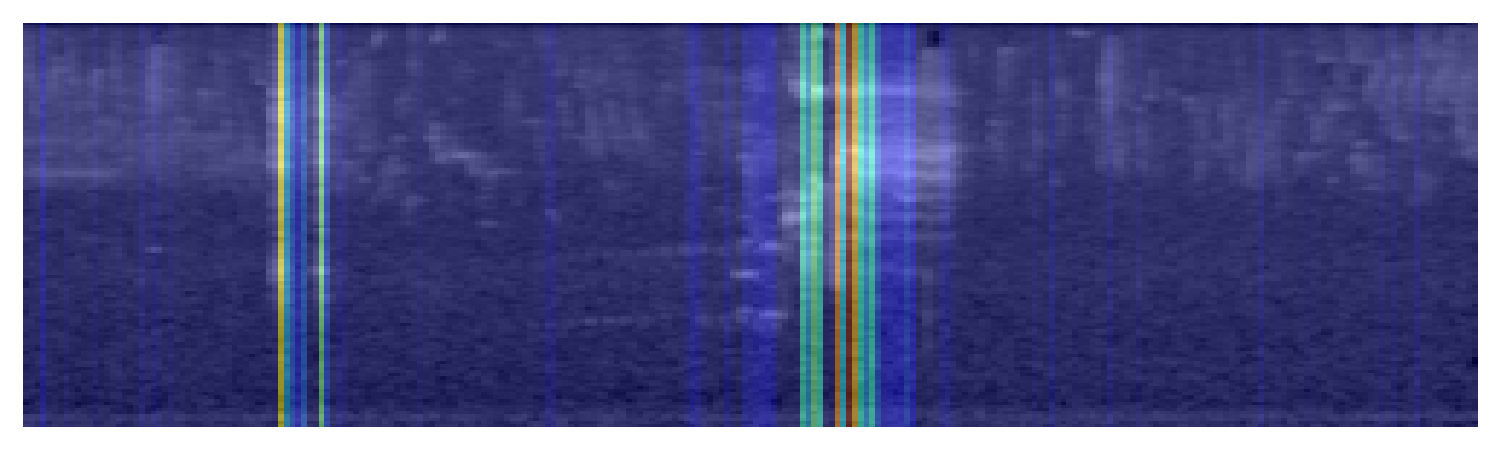}
    \includegraphics[width=\linewidth]{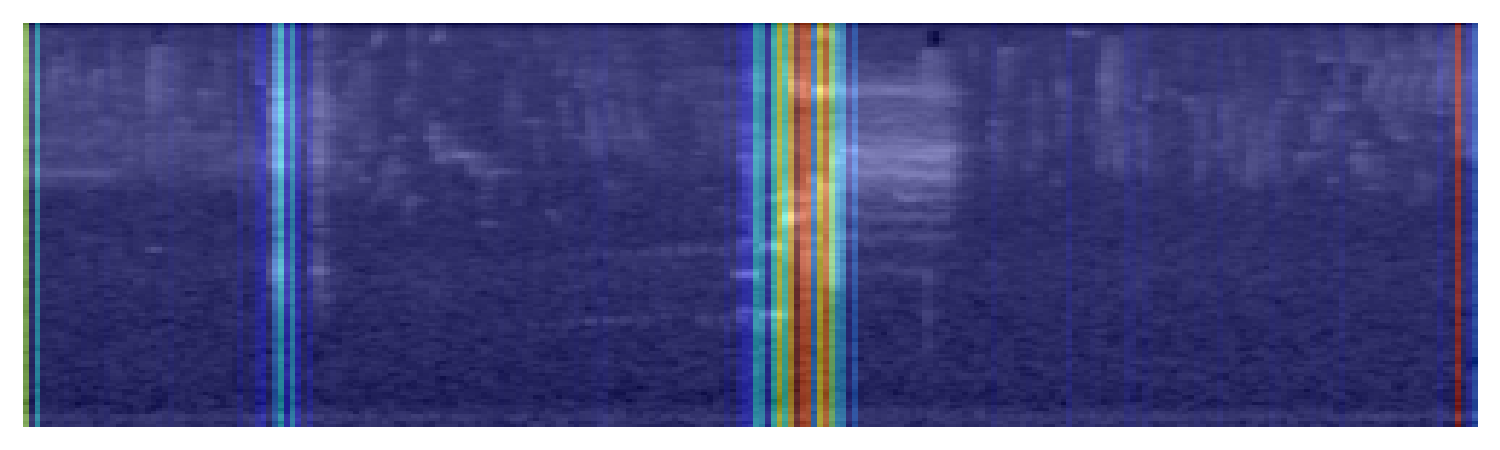}
    \includegraphics[width=\linewidth]{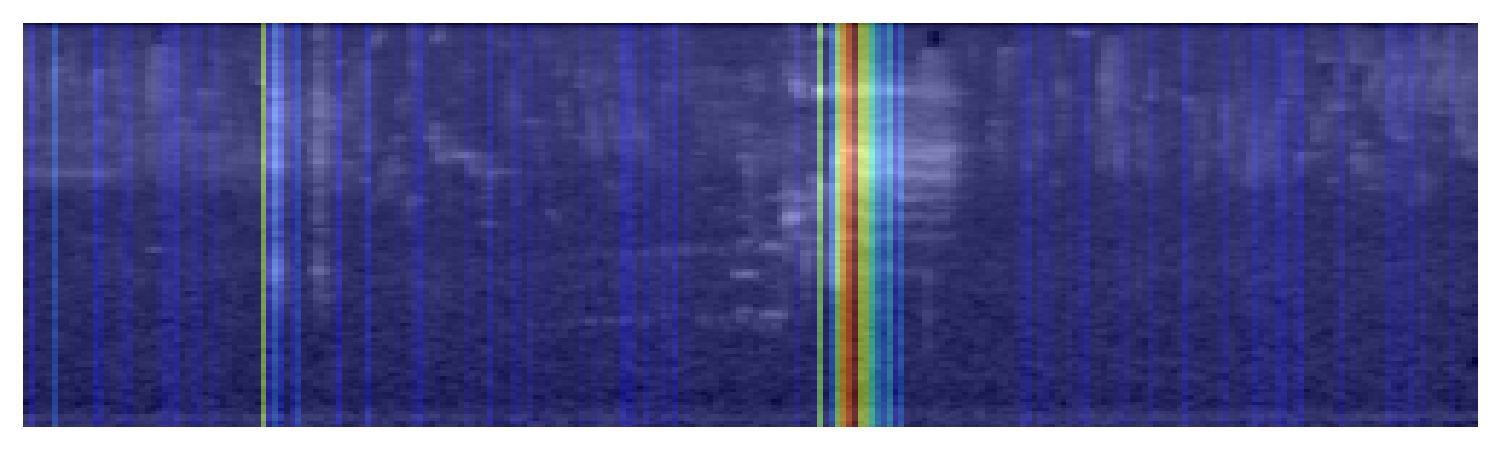}
    \caption*{Grad-CAM (rows: \cmark, \cmark, \cmark)}
\end{subfigure}%
\hfill
\begin{subfigure}{.49\linewidth}
    \includegraphics[width=\linewidth]{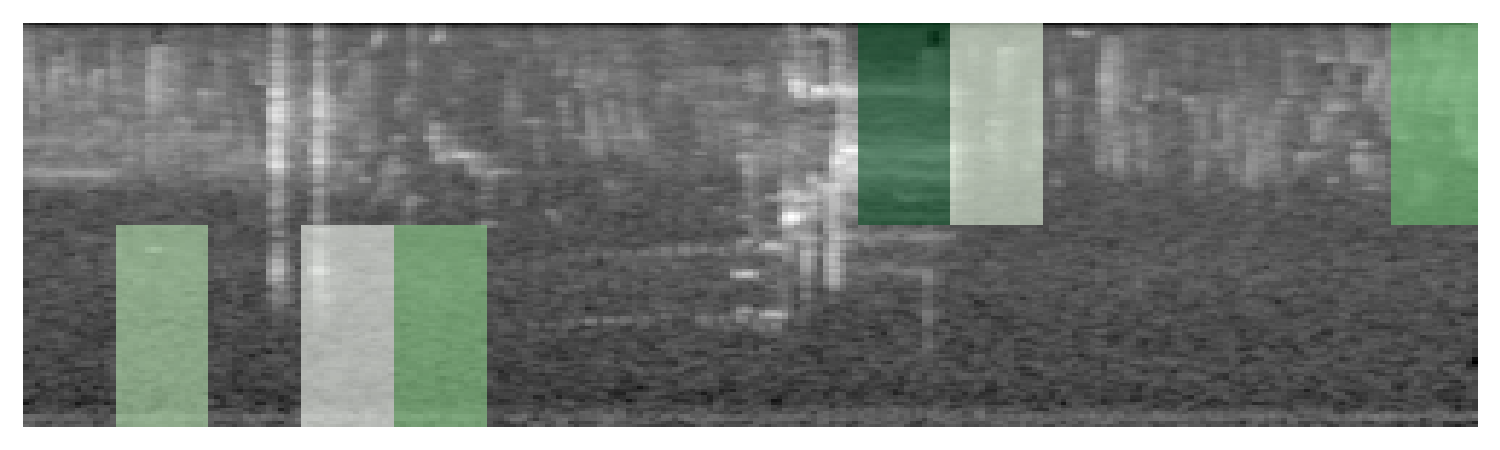}
    \includegraphics[width=\linewidth]{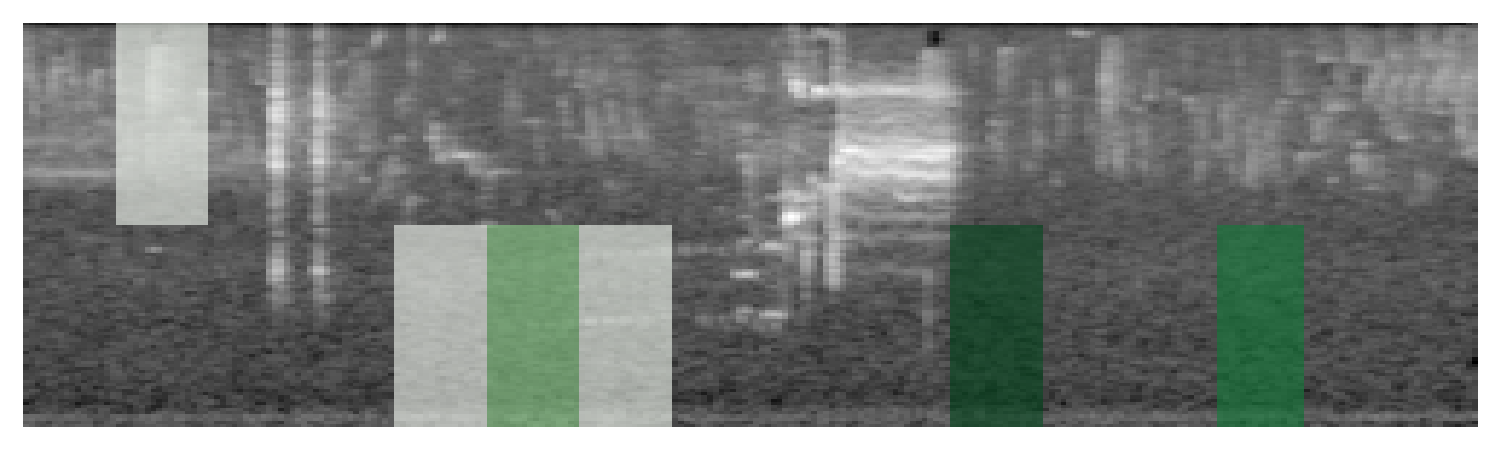}
    \includegraphics[width=\linewidth]{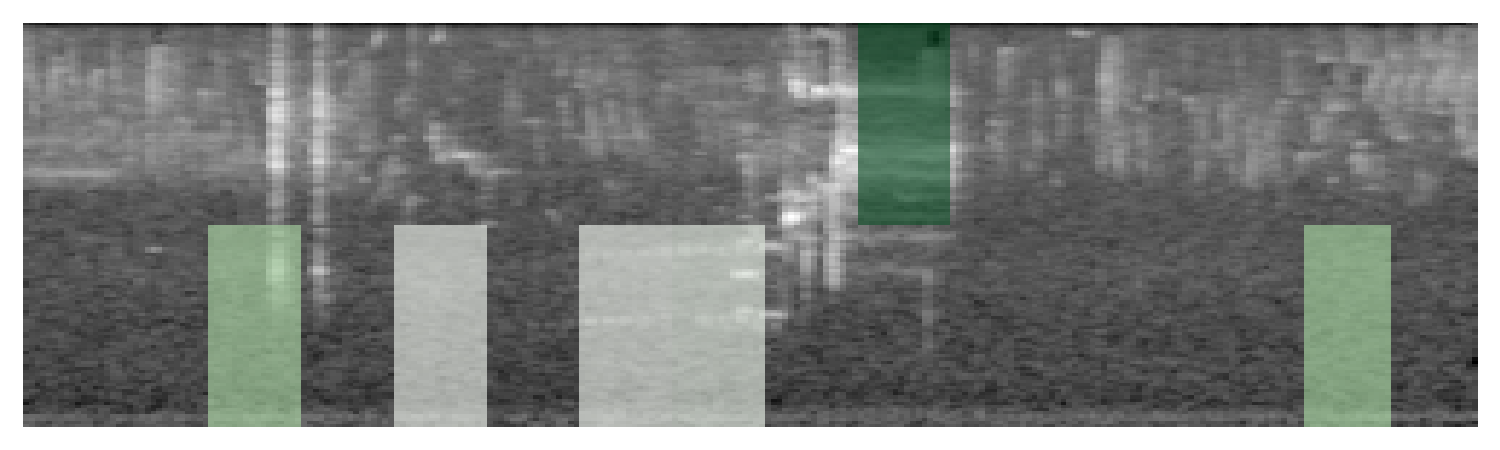}
    \caption*{LIME (rows: \cmark, \cmark, \cmark)}
\end{subfigure}

\caption{Correctly classified sample (\textit{A.\,phoeniceus}).}
\label{fig:case2}

\end{figure}

Figures \ref{fig:case1}–\ref{fig:case2} contrast a failure and a success scenario.

\noindent\textbf{Failure (dialect mismatch).}  
The D1-trained model (top row) aligns tightly with the tonal bursts, indicating reliable species cues.
The D2 model spreads attention across time, failing to isolate any spectral peak, whereas the D3 model attends confidently but off-target, illustrating an over-fit to spurious patterns.
Both explanations agree that only the region-matched network attends to dialect-invariant information, illustrating how dialect mismatch degrades generalisation.

\noindent\textbf{Success (dialect-invariant cues).}  
Figure \ref{fig:case2} shows the explainability results for an easier token of the same species.
Grad-CAM is almost identical across rows: each network locks onto the same vertical bands where the bird's tonal bursts occur.  
LIME is less crisp but still overlaps the active bands, indicating shared evidence even with different perturbation paths.


\section{Conclusion}
\label{sec:conclusion}
We presented a TDNN architecture that couples instance-frequency normalisation with gradient-reversal training and a two-stage augmentation pipeline. Standard perturbations and Mixup alleviate within-region data scarcity, while CycleGAN-based style transfer synthesizes missing dialect variants from Region 2, with DCA softly down-weighting synthetic samples to avoid overfitting. On DB3V, the model raises worst-case cross-region accuracy by more than twenty percentage points and lifts the low-resource D2 split by 11\%. Grad-CAM and LIME confirm that the network concentrates on stable harmonic bands, supporting ecological interpretation. Taken together, the results show that frequency-aware normalisation, targeted augmentation and light adversarial regularisation can deliver a deployable, interpretable solution for large-scale bird-sound monitoring across dialects. Future work will attempt other model architectures, end-to-end pipeline, and quantified explainability.


\bibliographystyle{IEEEtran}
\bibliography{refs}

\end{document}